 %%%%%%%This requires the PHYZZX.TEX macropackage

\tolerance=10000
\input phyzzx

%%%%%%%If you do not have the msbm fonts, delete the following 4 lines
\font\mybb=msbm10 at 12pt
\def\bbbb#1{\hbox{\mybb#1}}
\def\Z {\bbbb{Z}}
\def\R {\bbbb{R}}
%%%%%%%%%%%%
%%%and replace with the following 2 lines (without %)
%\def\Z {Z}
%\def\R {R}
%%%%%%%%%%

%%%%%%%%%%%%%%%%%%%%%%%%%%%%%%%%%

%%%%%%%%%%%%
%%%%%%%%%%%%
%\makeatletter
%\new@fontshape{msb}{m}{n}{%
%   <5>msbm5%
%%   <6>msbm6%
 %  <7>msbm7%
 %  <8>msbm8%
 %  <9>msbm9%
 %  <10>msbm10%
 %  <11>msbm10%
 %  <12>msbm10 at10.95pt%
 %  <14>msbm10 at14.4pt%
 %  <17>msbm10 at17.28pt%
 %  <20>msbm10 at20.74pt%
 %  <25>msbm10 at24.88pt}{}
%\extra@def{msb}{}{}%{\noaccents@}
%\newmathalphabet*\bbl{msb}{m}{n}
%\makeatother
%%\def\semi{{\bbl n}}%needsi Latex ? oldlfont??

\def\lsemidir{\mathbin{\hbox{\hskip2pt\vrule height 5.7pt depth -.3pt
width .25pt\hskip-2pt$\times$}}}

\def\sd{{\lsemidir}} %% semi-direct product

\def \mm {\mu}
\def \nn {\nu}

\def \ti {\tilde}

\def \2 {{1 \over 2}}
\def \3 {{1 \over 3}}
\def \4 {{1 \over 4}}
\def \5 {{1 \over 5}}
\def \6 {{1 \over 6}}
\def \7 {{1 \over 7}}
\def \8 {{1 \over 8}}
\def \9 {{1 \over 9}}
\def \0 { \infty}

\def\++ {{(+)}}
\def \- {{(-)}}
\def\+-{{(\pm)}}

%%%%%%%%%%%%%%%%%%%%%%%%%%%%%%%%%%%%%%%%%%%%%%%%%%%%%%%%%%%%%%%%%%%%

 \def\unit{\hbox to 3.3pt{\hskip1.3pt \vrule height 7pt width .4pt \hskip.7pt
\vrule height 7.85pt width .4pt \kern-2.4pt
\hrulefill \kern-3pt
\raise 4pt\hbox{\char'40}}}
\def\II{{\unit}}

\def\nup#1({Nucl.\ Phys.\  {\bf B#1}\ (}

%%%%%%%%%%%%%%%%%%%%%%%%%%%%%%%%%%%%%%%%%%%%%%%%%%%%%%%%%%%%%%%%%%%%
%%%%%%%%%%%%%%%%%%%%%%%%%%%%%%%%%%%%%%%%%%%%%%%%%%%%%%%%%%%%%%%%%%%%
\REF\CJ{E. Cremmer and B. Julia, Phys. Lett. {\bf 80B} (1978) 48; Nucl.
Phys. {\bf B159} (1979) 141.}
\REF\julia{B. Julia in {\it Supergravity and Superspace}, S.W. Hawking and M.
Ro$\check c$ek, C.U.P.
Cambridge,  (1981). }
\REF\cjlp{E. Cremmer, B. Julia, H. Lu and C. Pope, hep-th/9710119.}
\REF\jujh{B. Julia, {\it Infinite Lie algebras in Physics}
in {\it Johns Hopkins workshop on Unified field  theories
and beyond}, ed.  G. Domokos et al. Baltimore,  (June 1981).}
\REF\brma{P. Breitenlohner and D. Maison, Ann. Inst. H. Poincare, {\bf 46}
(1987) 215.}
\REF\Senb{A. Sen, Nucl. Phys. {\bf B447} (1995) 62, hep-th/9503057.}
\REF\junis{B. Julia and H. Nicolai, hep-th/9608082, Nucl. Phys.
{\bf B482} (1996) 431.}
\REF\beju{D. Bernard and B. Julia, Twisted Self-Duality of Dimensionally
Reduced Gravity and Vertex Operators, hep-th/9712254.}
\REF\juliab{B. Julia (1982) in {\it Lectures in Applied Mathematics},
AMS-SIAM,  {\bf 21} (1985) 355.}
\REF\HT{C.M. Hull and P.K. Townsend, hep-th/9410167.}
\REF\time{R. Geroch, J. Math. Phys. {\bf 12} (1971) 918; W. Kinnersley, J.
Math.  Phys. {\bf 14} (1973) 651.}
\REF\Bob{B.S. Acharya, M. O'Loughlin and B. Spence, Nucl.
Phys. {\bf B503} (1997) 657; B.S. Acharya, J.M. Figueroa-O'Farrill, M.
O'Loughlin and B. Spence,
hep-th/9707118.}
\REF\thom{M. Blau and G. Thompson, Phys. Lett. {\bf B415} (1997) 242.}
\REF\hagi{G. Gibbons and S. Hawking, Com. Math. Phys. {\bf 66} (1979) 291.}
\REF\gib{G. Gibbons, P. Breitenlohner and D. Maison, Com. Math. Phys. {\bf 120}
(1988) 295.}
\REF\moore{G. Moore, hep-th/9305139, hep-th/9308052.}
\REF\junic{B. Julia  and H. Nicolai, Nucl. Phys. {\bf B439} (1995) 291}
\REF\Mat{T. Banks, W. Fischler, S. Shenker and L. Susskind,
``M theory as a matrix model: a conjecture,''
hep-th/9610043, Phys. Rev.
{\bf D55} (1997) 5112.}%
\REF\brev{For  reviews see T. Banks,
hep-th/9706168, hep-th/9710231.}%
\REF\Sus{L. Susskind, hep-th/9704080.}
\REF\Sen{A. Sen, hep-th/9709220.}
\REF\Seib{N. Seiberg,
hep-th/9710009.}%
\REF\conc{A. Connes, Noncommutative Geometry, Academic Press, (1994)}
\REF\dwe{B. de Wit, H.J. Matschull and H. Nicolai Phys. Lett.  {\bf B318}
(1993)
 115}
\REF\ehlers{J. Ehlers, Dissertation, Hamburg (1957)}
\REF\TD {A. Giveon, M. Porrati and E. Rabinovici, Phys. Rep. {\bf 244}
(1994) 77.}
\REF\Cham {A.H. Chamseddine, Nucl. Phys. {\bf B185} (1981) 403.}
\REF\Narain{K.S.  Narain, Phys. Lett. {\bf B169} (1986) 41.}
\REF\NSW {K.S.  Narain,  M.H. Sarmadi and E. Witten, Nucl. Phys. {\bf B279}
(1987) 369.}
\REF\SaS{A. Salam and E. Sezgin, Supergravities in various dimensions vol.1,
North Holland/World Scientific, (1989)}
\REF\Sena{A. Sen, Nucl. Phys. {\bf B434} (1995) 179, hep-th/9408083.}

%%%%%%%%%%%%%%%%%%%%%%%%%%%%%%%%%%%%%%%%%%%%%%%%%%%%%%%%%%%%%%%%%%%%

\Pubnum{ \vbox{  \hbox {QMW-PH-98-15} \hbox{LPTENS 98/09}
\hbox{hep-th/9803239}} }
\pubtype{}
\date{March, 1998}

\titlepage

\title {\bf  Duality and Moduli Spaces for Time-Like
 Reductions}

\author{C.M. Hull}
\address{Physics Department, Queen Mary and Westfield College,
\break Mile End Road, London E1 4NS, U.K.}
%\andaddress{Laboratoire de Physique Th\' eorique, Ecole Normale Sup\' erieure,
%24 Rue Lhomond, 75231 Paris Cedex 05, France.}
\vskip 0.5cm

\andauthor{B. Julia}
\address{Laboratoire de Physique Th\' eorique, Ecole Normale Sup\' erieure,
24 Rue Lhomond, 75231 Paris Cedex 05, France.}

\abstract {We consider the dimensional reduction/compactification
of supergravity, string and M-theories on
tori with one time-like circle. We find the coset spaces in which the massless
scalars take their
values, and identify the discrete duality groups. }

\endpage

%%%%%%%%%%%%%%%%%%%%%%%%%%%%%%%%
%
% S-Tables Macro
%
%\message{S-Tables Macro v1.0, ACS, TAMU (RANHELP@VENUS.TAMU.EDU)}
%
% Help Text
%
\newhelp\stablestylehelp{You must choose a style between 0 and 3.}%
\newhelp\stablelinehelp{You should not use special hrules when stretching
a table.}%
\newhelp\stablesmultiplehelp{You have tried to place an S-Table inside another
S-Table.  I would recommend not going on.}%
%
% Line Thicknesses (Values)
%
\newdimen\stablesthinline
\stablesthinline=0.4pt
\newdimen\stablesthickline
\stablesthickline=1pt
%
% Border and Internal Line Thicknesses
%
\newif\ifstablesborderthin
\stablesborderthinfalse
\newif\ifstablesinternalthin
\stablesinternalthintrue
\newif\ifstablesomit
\newif\ifstablemode
\newif\ifstablesright
\stablesrightfalse
%
% Save Registers
%
\newdimen\stablesbaselineskip
\newdimen\stableslineskip
\newdimen\stableslineskiplimit
%
% Counters
%
\newcount\stablesmode
\newcount\stableslines
\newcount\stablestemp
\stablestemp=3
\newcount\stablescount
\stablescount=0
\newcount\stableslinet
\stableslinet=0
%
% Table Style Selection
%
% 0 - Centered
% 1 - Left Justified
% 2 - Right Justified
% 3 - Not Justified
%
\newcount\stablestyle
\stablestyle=0
%
% Element Buffering Definitions
%
\def\stablesleft{\quad\hfil}%
\def\stablesright{\hfil\quad}%
%
% Vertical Bar Activation
%
\catcode`\|=\active%
%
% Strut Control
%
\newcount\stablestrutsize
\newbox\stablestrutbox
\setbox\stablestrutbox=\hbox{\vrule height10pt depth5pt width0pt}
\def\stablestrut{\relax\ifmmode%
                         \copy\stablestrutbox%
                       \else%
                         \unhcopy\stablestrutbox%
                       \fi}%
%
% Misc. Internal Stuff
%
\newdimen\stablesborderwidth
\newdimen\stablesinternalwidth
\newdimen\stablesdummy
\newcount\stablesdummyc
\newif\ifstablesin
\stablesinfalse
%
% Table Macros
%
\def\begintable{\stablestart%
  \stablemodetrue%
  \stablesadj%
  \halign%
  \stablesdef}%
\def\stablesadj{%
  \ifcase\stablestyle%
    \hbox to \hsize\bgroup\hss\vbox\bgroup%
  \or%
    \hbox to \hsize\bgroup\vbox\bgroup%
  \or%
    \hbox to \hsize\bgroup\hss\vbox\bgroup%
  \or%
    \hbox\bgroup\vbox\bgroup%
  \else%
    \errhelp=\stablestylehelp%
    \errmessage{Invalid style selected, using default}%
    \hbox to \hsize\bgroup\hss\vbox\bgroup%
  \fi}%
\def\stablesend{\egroup%
  \ifcase\stablestyle%
    \hss\egroup%
  \or%
    \hss\egroup%
  \or%
    \egroup%
  \or%
    \egroup%
  \else%
    \hss\egroup%
  \fi}%
\def\stablestart{%
  \ifstablesin%
    \errhelp=\stablesmultiplehelp%
    \errmessage{An S-Table cannot be placed within an S-Table!}%
  \fi
  \global\stablesintrue%
  \global\advance\stablescount by 1%
  \message{<S-Tables Generating Table \number\stablescount}%
  \begingroup%
  \stablestrutsize=\ht\stablestrutbox%
  \advance\stablestrutsize by \dp\stablestrutbox%
  \ifstablesborderthin%
    \stablesborderwidth=\stablesthinline%
  \else%
    \stablesborderwidth=\stablesthickline%
  \fi%
  \ifstablesinternalthin%
    \stablesinternalwidth=\stablesthinline%
  \else%
    \stablesinternalwidth=\stablesthickline%
  \fi%
  \tabskip=0pt%
  \stablesbaselineskip=\baselineskip%
  \stableslineskip=\lineskip%
  \stableslineskiplimit=\lineskiplimit%
  \offinterlineskip%
  \def\borderrule{\vrule width \stablesborderwidth}%
  \def\internalrule{\vrule width \stablesinternalwidth}%
  \def\thinline{\noalign{\hrule height \stablesthinline}}%
  \def\thickline{\noalign{\hrule height \stablesthickline}}%
  \def\trule{\omit\leaders\hrule height \stablesthinline\hfill}%
  \def\ttrule{\omit\leaders\hrule height \stablesthickline\hfill}%
  \def\tttrule##1{\omit\leaders\hrule height ##1\hfill}%
  \def\stablesel{&\omit\global\stablesmode=0%
    \global\advance\stableslines by 1\borderrule\hfil\cr}%
  \def\el{\stablesel&}%
  \def\elt{\stablesel\thinline&}%
  \def\eltt{\stablesel\thickline&}%
  \def\elttt##1{\stablesel\noalign{\hrule height ##1}&}%
  \def\elspec{&\omit\hfil\borderrule\cr\omit\borderrule&%
              \ifstablemode%
              \else%
                \errhelp=\stablelinehelp%
                \errmessage{Special ruling will not display properly}%
              \fi}%
  \def\stmultispan##1{\mscount=##1 \loop\ifnum\mscount>3 \stspan\repeat}%
  \def\stspan{\span\omit \advance\mscount by -1}%
  \def\multicolumn##1{\omit\multiply\stablestemp by ##1%
     \stmultispan{\stablestemp}%
     \advance\stablesmode by ##1%
     \advance\stablesmode by -1%
     \stablestemp=3}%
  \def\multirow##1{\stablesdummyc=##1\parindent=0pt\setbox0\hbox\bgroup%
    \aftergroup\emultirow\let\temp=}
  \def\emultirow{\setbox1\vbox to\stablesdummyc\stablestrutsize%
    {\hsize\wd0\vfil\box0\vfil}%
    \ht1=\ht\stablestrutbox%
    \dp1=\dp\stablestrutbox%
    \box1}%
  \def\stpar##1{\vtop\bgroup\hsize ##1%
     \baselineskip=\stablesbaselineskip%
     \lineskip=\stableslineskip%
     \lineskiplimit=\stableslineskiplimit\bgroup\aftergroup\estpar\let\temp=}%
  \def\estpar{\vskip 6pt\egroup}%
  \def\stparrow##1##2{\stablesdummy=##2%
     \setbox0=\vtop to ##1\stablestrutsize\bgroup%
     \hsize\stablesdummy%
     \baselineskip=\stablesbaselineskip%
     \lineskip=\stableslineskip%
     \lineskiplimit=\stableslineskiplimit%
     \bgroup\vfil\aftergroup\estparrow%
     \let\temp=}%
  \def\estparrow{\vfil\egroup%
     \ht0=\ht\stablestrutbox%
     \dp0=\dp\stablestrutbox%
     \wd0=\stablesdummy%
     \box0}%
  \def|{\global\advance\stablesmode by 1&&&}%
  \def\|{\global\advance\stablesmode by 1&\omit\vrule width 0pt%
         \hfil&&}%
  \def\vt{\global\advance\stablesmode by 1&\omit\vrule width \stablesthinline%
          \hfil&&}%
  \def\vtt{\global\advance\stablesmode by 1&\omit\vrule width
\stablesthickline%
          \hfil&&}%
  \def\vttt##1{\global\advance\stablesmode by 1&\omit\vrule width ##1%
          \hfil&&}%
  \def\vtr{\global\advance\stablesmode by 1&\omit\hfil\vrule width%
           \stablesthinline&&}%
  \def\vttr{\global\advance\stablesmode by 1&\omit\hfil\vrule width%
            \stablesthickline&&}%
  \def\vtttr##1{\global\advance\stablesmode by 1&\omit\hfil\vrule width ##1&&}%
  \stableslines=0%
  \stablesomitfalse}
\def\stablesdef{\bgroup\stablestrut\borderrule##\tabskip=0pt plus 1fil%
  &\stablesleft##\stablesright%
  &##\ifstablesright\hfill\fi\internalrule\ifstablesright\else\hfill\fi%
  \tabskip 0pt&&##\hfil\tabskip=0pt plus 1fil%
  &\stablesleft##\stablesright%
  &##\ifstablesright\hfill\fi\internalrule\ifstablesright\else\hfill\fi%
  \tabskip=0pt\cr%
  \ifstablesborderthin%
    \thinline%
  \else%
    \thickline%
  \fi&%
}%
\def\endtable{\advance\stableslines by 1\advance\stablesmode by 1%
   \message{- Rows: \number\stableslines, Columns:  \number\stablesmode>}%
   \stablesel%
   \ifstablesborderthin%
     \thinline%
   \else%
     \thickline%
   \fi%
   \egroup\stablesend%
\endgroup%
\global\stablesinfalse}
%
% end of STABLES.TEX
%

%\chapter{Introduction}

The standard dimensional reduction of 11-dimensional supergravity [\CJ,\julia]
on a torus $T^d$
gives a supergravity theory in $11-d$ dimensions which is invariant under a
rigid duality symmetry
$G_d$ and a local symmetry $H_d$; the groups $G_d, H_d$ are listed in table 1.
$H_d$
is the maximal compact subgroup of $G_d$ and the theory has scalars taking
values in the coset
$G_d/H_d$.
If  in $D$ dimensions some of the $p$-form gauge fields for various values of
$p$ are dualised to
$(D-p-2)$-form gauge fields, then
 Noether and topological charges are interchanged, and
 internal symmetries are traded for (generalized) gauge symmetries  [\cjlp].
The reduction to two dimensions is particularly
involved but follows  the higher dimensional pattern, see for instance
[\jujh -\beju].  In  that case the duality groups  are
infinite dimensional, with affine Kac-Moody symmetries and  the associated
Virasoro (Witt) symmetries.
In one dimension, it is conjectured that there is an $E_{10(10)}$ symmetry
[\juliab].

The duality $G_d$ of the dimensionally reduced supergravity theory was first
discovered as a symmetry of the compactified theory after discarding the
nonzero Kaluza-Klein modes.
In any
quantum theory for which the supergravity theory is a low-energy effective
description, the
rigid $G_d$ invariance is broken to (at most) a discrete subgroup
$G_d(\Z)$ [\HT]. It is conjectured that
11-dimensional supergravity is a low-energy effective description of a
consistent quantum M-theory,
and that $G_d(\Z)$ is a symmetry of the toroidally compactified theory,
the U-duality symmetry [\HT]. (We will distinguish between dimensional
reduction, in which the Kaluza-Klein modes are
truncated, and compactification, in which they are kept.)

\vskip 1cm

\begintable
D=11-d |
$G_d$ | $H_d$  |  {U-Duality}  \elt
 $10$ | $SO(1,1) $ | $\II $ | - \elt
 $9$ | $SL(2,\R)\times SO(1,1)$ | $SO(2) $ | $SL(2,\Z) $ \elt
 $8$ | $SL(3,\R)\times SL(2,\R)$ |  $SO(3)\times SO(2) $ | $SL(3,\Z)\times
SL(2,\Z)$
\elt
 $7$ | $SL(5,\R)$ | $SO(5)$ | $SL(5,\Z)$ \elt
 $6$ | $SO(5,5)$ | $SO(5)\times SO(5)$ |  $SO(5,5;\Z)$ \elt
 $5$ | $E_{6(6)}$ | $USp(8)$ | $E_{6(6)}(\Z)$ \elt
 $4$ | $E_{7(7)}$ | $SU(8)$ | $E_{7(7)}(\Z)$ \elt
 $3$ | $E_{8(8)}$ | $SO(16)$ | $E_{8(8)}(\Z)$
\elt
 $2$ | $E_{9(9)}$ | $H_2$ | $E_{9(9)}(\Z)$
%\elt
% $1$ | $E_{10(10)}$? | -- |  --
\endtable

%\centerline
{{\bf Table 1} Duality symmetries for 11 dimensional supergravity reduced, and
 M-theory compactified to $D=11-d$ dimensions on $T^d$. The classical scalar
symmetric space is $G_d/H_d$ and the U-duality group is $G_d(\Z)$.}

\vskip .5cm
In three dimensions, the bosonic sector consists of gravity coupled to
an $E_8/SO(16)$ coset space, and reduction to $D=2$ gives scalars in
$E_8/SO(16)\times \R$, coupled to gravity and a vector field.
This has an infinite dimensional symmetry [\julia]
  $E_{9(9)}$ (which is  an affine  $E_{8(8)}$), and
the quantum U-duality symmetry is conjectured to be a discrete subgroup of this
[\HT].
 The natural  subgroup  $H_2$    to consider in this context   is an infinite
dimensional subgroup of  $E_{9(9)}$ containing
$SO(16)$,
which can be characterised as a  fixed set of a Cartan involution of
$E_{9(9)}$,
as in the case of symmetric spaces.   It is also the subspace of the adjoint
representation on which the Killing form, or rather the invariant bilinear
form, is negative definite [\jujh] and deserves the
name of maximal compact subalgebra by analogy with the finite dimensional
semi-simple situation with  the Cartan compactness criterion. $H_2$, as acting
on solutions in the Geroch [\time] group fashion, does not contain the central
charge of $E_9$.  A different subalgebra arises in the dressing method, see
[\beju].

Instead of reducing on a space-like torus to obtain a theory with Lorentzian
signature in $11-d$
dimensions, one can instead compactify time along with $d-1$ spatial directions
and reduce on the
Lorentzian-signature torus $T^{d-1,1}$ to obtain a  supergravity theory in
$11-d$
dimensions with Euclidean-signature. Truncating  the dependence on the internal
dimensions leaves a dimensionally reduced
theory in which some of the fields have kinetic terms of the wrong sign, but in
the full compactified theory
such ghosts can be gauged away. This is readily seen in the example of
dimensional reduction of (super) Yang-Mills theory from
$D+d$ dimensions    to $D$ dimensions. Reducing on a Euclidean torus $T^d$
gives Yang-Mills  plus $d$ scalars in Minkowski
space, acted on by an $SO(d)$ \lq R-symmetry', while reducing on a Lorentzian
torus $T^{d-1,1}$ gives Yang-Mills
 plus $d$ scalars in Euclidean
space, acted on by an $SO(d-1,1)$ \lq R-symmetry'. In particular, the scalar
field coming from $A_0$, the time component of
the gauge field, has a kinetic term of the wrong sign. However, in the full
compactified theory, one can
eliminate $A_0$ by a gauge choice (e.g.
$A_0=0$)  leaving a theory without ghosts.
Such dimensional reductions of super Yang-Mills are
 useful in constructing topological field theories [\Bob,\thom].

Timelike dimensional reductions    of  supergravity and string theories were
investigated in
[\hagi, \juliab, \gib, \moore],
and can be
useful for
constructing stationary solutions in $ 12-d$ dimensions. For example, in [\gib]
the case $d=8$ was considered and the resulting Euclidean 3-dimensional theory
was
used to construct stationary solutions of $N=8$ supergravity in $3+1$
dimensions.
In [\gib], it was found that the Euclidean maximal supergravity in 3 dimensions
still had
$G_d=E_{8(+8)}$ but that $H_d$, which had been $SO(16)$ in $2+1$ dimensions,
had become
the non-compact group $SO^*(16)$; the noncompact $SO^*(2k)$ had been discussed
already in [\juliab].
In [\moore], the bosonic string was compactified to zero dimensions on
$T^{25,1}$, allowing an
algebraic study of the theory.

The remaining possibility is to consider compactification on a torus which
includes a
light-like circle; such
null
reductions were studied in [\junic].
This could be viewed as the result of compactifying on a spacelike torus,
boosting in one of the toroidal directions and taking the limit of infinite
boost. M-theory
compactifications on such null tori have recently been the focus of
considerable
interest, as taking
a suitable limit defines the matrix theory for toroidal
compactifications of
M-theory [\Mat-\Seib]. It is of particular interest to study the duality
symmetries for such null reductions and
compactifications, and their implications for the U-duality of M-theory. Such
reductions would also
be useful in constructing solutions with null Killing vectors.

Our aim here is to study the duality symmetries and scalar coset spaces of
gravity and supergravity
theories compactified on tori including a time-like circle, and to return to
the null case in a future publication.

For compactification of gravity on a space $X$, the dimensionally reduced
theory has massless
scalars taking values in the moduli space of Ricci-flat metrics on $X$,
$M_X$.
For supergravity theories, or for gravity coupled to anti-symmetric tensor
gauge fields, there are
moduli for both gravity and the anti-symmetric tensor gauge fields.
For compactification on a Ricci-flat $X$, with all other fields  vanishing,
there are, in the linear approximation,
massless scalars  corresponding to flat anti-symmetric tensor gauge
fields on $X$ that extend
the moduli space of Ricci-flat metrics $M_X$ into some moduli space $N_X$ in
which all massless scalars take their values.
There will be massless scalars taking values in $N_X$ independent of whether
one keeps the full
Kaluza-Klein theory (compactification) or whether one performs a consistent
truncation to the massless sector (dimensional reduction).
For compactification of string or M-theories,
the massless fields can be studied using the appropriate (super)-gravity
theory, and they again
take values in the appropriate $N_X$.

In this paper we will discuss compactifications in which $X$ is a torus and the
moduli space takes
the form
$M_G=G(\Z) \backslash G/H$ for some groups $G,H$ where $G(\Z)$ is a discrete
subgroup of $G$.
For compactification on Euclidean tori, $H$ is the maximal compact subgroup of
$G$
so that $G/H$ is a Riemannian manifold, a symmetric space of the noncompact
type, and $M_G$ is a
Hausdorff space with cusps in which the massless scalars take their values.
For compactification on a torus with a time-like direction, the situation is
more complicated
[\moore].
In this case, $H$ is non-compact and $G/H$ has indefinite signature.
%The scalar fields arising from the dimensional reduction of supergravity,
%which are parameterised by
%metrics and anti-symmetric tensors on $X$, take values in a
%{\it proper open subset } of $G(\Z) \backslash G/H$, and as a result, t
%The action of $G(\Z)$ on the supergravity data is not always well-defined, o
%One may, according to G. Moore, need a larger set of data than supergravity
% data
% to describe M-theory vacua.
The action of $H$ on
$G(\Z) \backslash G $ is typically {\it ergodic}, with generic orbits being
space-filling.
This leads to  non-commutative geometry [\moore, \conc].
(See also [\dwe].)
In the following we will not discuss such points further, but restrict
ourselves
to finding the groups $G,H,G(\Z)$ that arise for various compactifications.

\noindent {\caps\enspace  Toroidal Compactifications of Gravity}
\vskip\headskip

Consider the dimensional reduction of $D+d$-dimensional gravity on $T^d$ to
$D$
dimensions.
The $(D+d)$-dimensional metric $G_{MN}$
gives a
$D $-dimensional metric $g_{\mm \nn}$, $d$ vector fields $g_{i\mm}$ and
$d(d+1)/2$ scalars, represented by a metric $g_{ij}$ on $T^d$.
For a Euclidean torus, the reduced theory has a rigid $GL(d,\R)$ symmetry and
the scalars can be
viewed as taking values in $GL(d,\R)/SO(d)$, which is the moduli space of flat
metrics on  $T^d$.
There is a subgroup $GL(d,\Z)$ which preserves the periodicities of the
toroidal directions and
which is a discrete gauge symmetry of the compactified theory, so that
configurations related by
$GL(d,\Z)$ transformations should be identified.

If instead one reduces on a torus $T^{d-1,1}$ with $d-1$ space-like circles and
  one time-like circle, a similar theory results, with rigid $GL(d,\R)$
symmetry but with scalars
taking values in $GL(d,\R)/SO(d-1,1)$. As  $GL(d,\R)$ is a remnant of the
$(D+d)$-dimensional
diffeomorphisms [\CJ, \julia], it remains a symmetry whatever the metric on
the torus, and the discrete subgroup $GL(d,\Z)$
remains as the discrete gauge symmetry.
However, the scalar coset space is now the moduli space of
Lorentzian metrics on $T^d$, which is $GL(d,\R)/SO(d-1,1)$ (as $SO(d-1,1)$ is
the subgroup of
 $GL(d,\R)$ preserving a Lorentzian metric). The scalar coset space is no
longer
a Riemannian space.

For reductions to $D=3$ dimensions, the $d$ vector fields $g_{i\mm}$ can be
dualised to give additional scalars,
and the $GL(d,\R)$ symmetry is enlarged by
 Ehlers-type transformations.
For example, reduction from $3+1$ Lorentzian dimensions on a  circle
gives a scalar and a vector, which can be dualised to give a second scalar. For
reductions on a spacelike circle,
the two scalars take values in $SL(2,\R) / SO(2)$ and  the Ehlers symmetry is
$SL(2,\R)$ [\ehlers, \time].
For reduction on a timelike circle, there are two minus signs which conspire to
leave the structure intact, so that the
scalar coset is again $SL(2,\R) / SO(2)$, but
$ 4$-dimensional  euclidean gravity  reduced to 3 dimensions leads to
the coset space $SL(2,\R) / SO(1,1)$  [\hagi].

\noindent {\caps\enspace  Bosonic and Heterotic String  Compactifications}
\vskip\headskip

Consider first the bosonic string compactified on $T^d$ (with $D+d=26$), or
a theory of $(D+d)$-dimensional gravity coupled to a 2-form gauge field and
dilaton
with action
$$
\int d^{D+d}x \sqrt{-g} e^{-2\Phi}\left(
R+ {1\over 12}  H^2 + 4 (\nabla \Phi)^2
\right)
\eqn\abc$$
We will first consider the case of reduction to more than 4 dimensions,
 with $D>4$.
 For a Euclidean torus,
this has a moduli space
$$\R\times
{
O(d,d)\over
O(d)\times O(d) }
\eqn\euco$$
with $O(d,d;\Z)$ being the T-duality symmetry of the theory, and the $\R$
factor representing the
dilaton. The coset space $O(d,d)/
O(d)\times O(d)$ has dimension $d^2$ and is parameterised by constant metric
$g_{ij}$
and anti-symmetric tensor $b_{ij}$ on $T^d$.
For a Lorentzian torus $T^{d-1,1}$, we expect a coset space $G_d/H_d$ with the
following properties.
(i)  As
$T^{d-1,1}$ contains both $T^{d-1,0}$ and $T^{d-2,1}$, we must have
$${
O(d-1,d-1)\over
O(d-1)\times O(d-1) } \subset G_d/H_d, \qquad  G_{d-1}/H_{d-1}\subset G_d/H_d
$$
(ii)  As $
{
GL(d,\R)/
 SO(d-1,1) } $ is the moduli space of metrics on $T^{d-1,1}$, we must have
$$
{
GL(d,\R)\over
 SO(d-1,1) } \subset G_d/H_d$$
(iii) The dimension of $ G_d/H_d$ is $d^2$, and is parameterised
by constant Lorentzian metrics   $g_{ij}$ taking values in ${
GL(d,\R)/
 SO(d-1,1) } $
and an anti-symmetric tensor $b_{ij}$ on $T^{d-1,1}$.
Splitting the torus coordinates $x^i$ into a time coordinate $x^0$ and $d-1$
space coordinates
$x^a$,  the $d(d-1)/2$ degrees of freedom corresponding to $b_{ij}$ split into
$s= (d-1)(d-2)/2$ degrees of freedom $b_{ab}$ with space-like norm and
$d$ degrees of freedom $b_{a0}$ with time-like norm, so that

$$G_d/H_d \sim {(GL(d,\R) \sd \R^{s,d}) \over SO(d-1,1) } $$
The simplest coset space with these properties is $G_d=O(d,d)$ and
$H_d=O(d-1,1)\times O(d-1,1) $,
so that the full moduli space is
$$
{
O(d,d)\over
O(d-1,1)\times O(d-1,1) } \times \R
\eqn\modsp$$
and this is indeed the correct moduli space [\juliab, \TD, \moore].
The effective action has
a rigid $O(d,d)\times \R$ symmetry, and the discrete subgroup  $O(d,d;\Z)$ is
again the T-duality
symmetry of the perturbative string theory.

This is easily generalised to the heterotic (or type I) string toroidally
compactified
from $ 10 $ dimensions.
Compactification on $T^d$ gives the moduli space
[\Cham, \juliab, \Narain, \NSW]
$$
\R\times {
O(d,d+n)\over
O(d)\times O(d+n) }
\eqn\abc$$
with $n=16$, while
similar arguments to the above give the moduli spaces
$$
{
O(d,d+n)\over
O(d-1,1)\times O(d+n-1,1) }\times \R
\eqn\abc$$
 for compactification on $T^{d-1,1}$.
The same spaces emerge for toroidal compactification of half-maximal
supergravity coupled to $n$
vector multiplets, for any $n$ [\SaS].

\noindent {\caps\enspace  Compactifications to 4 Dimensions and Less}
\vskip\headskip

For reduction of the bosonic or heterotic string, or of the related
supergravity theories
to 4 dimensions or less gives scalars in the moduli spaces  \euco\  or
 \modsp, as before. However,
in 4 dimensions the 2-form gauge field $b_{\mm \nn}$ can be dualised to a
pseudo-scalar, which
combines with the dilaton in the $\R$ factor to take values in a
two-dimensional coset space .
For the usual compactification on a spacelike torus
this is
$SL(2,\R)/SO(2)$,
while for compactification on a Lorentzian torus
it is $SL(2,\R)/SO(1,1)$ (the difference in sign for the pseudoscalar kinetic
term coming from the duality transformation).
Thus the coset space for the heterotic string compactified on a Lorentzian
torus $T^{5,1}$ to Euclidean 4-space is
$$
{
O(6,22)\over
O(5,1)\times O(11,1) }\times
{SL(2,\R) \over
SO(1,1)}
\eqn\abc$$
In both the Lorentzian and the Euclidean cases, there is a  non-perturbative
$SL(2,\Z)$ S-duality
symmetry.

Note that a similar structure can occur in $N=4$ super Yang-Mills in 4
dimensions.
Starting from the (2,0) supersymmetric self-dual tensor multiplet theory in 6
dimensions, reducing on $T^2$ gives the usual Lorentzian $N=4$ super
Yang-Mills with coupling constants in
$SL(2,\R)/SO(2)$, while
compactifying on $T^{1,1}$ gives a Euclidean
$N=4$ super Yang-Mills with coupling constants in  $SL(2,\R)/SO(1,1)$.
In both cases, the moduli of the torus become the coupling constants on
reduction and the
$SL(2,\Z)$ torus symmetry becomes the S-duality of the 4-dimensional theory.
This Euclidean theory can be twisted to give a topological field theory, as in
[\Bob, \thom].

Toroidal compactification of gravity from $d+3$ dimensions to 3 dimensions
gives
scalars in
${
GL(d,\R)/
 SO(d)}$
for a space-like torus  $T^d$
or $
{
GL(d,\R)/
 SO(d-1,1) }$
for a Lorentzian torus $T^{d-1,1}$,
together with $d$ vector fields from the reduction of the metric.
In 3 dimensions, these can be dualised to give an extra $d$ scalars.
The dimensional reduction of gravity from 4 to 3 dimensions gives rise to an
$SL(2,\R)$ Ehlers
symmetry [\ehlers], and there are $d$ such    $SL(2,\R)$ symmetries,
corresponding to the ways of first
reducing on a $d-1$ torus to four dimensions, and then reducing from four to
three dimensions.
There are then these $d$ {}     $SL(2,\R)$ symmetries together with the
geometrical
 $GL(d,\R)$
symmetry, which do not commute but fit together to generate the group
$SL(d+1,\R)$
of symmetries of the 3-dimensional field theory.
For reduction on a spacelike torus, the scalars in ${
GL(d,\R)/
 SO(d)}$
and the extra $d$ scalars from dualising the vectors combine to take values in
the coset space
$$
{
SL(d+1,\R)\over
 SO(d+1)}$$
For a Lorentzian torus, the extra scalars transform as a {\bf d} of
${SO(d-1,1) }$
and so, after dualisation $(d-1)$ of those also  have a kinetic term of the
 wrong sign. These $d$ scalars  combine
with those in
 $
{
GL(d,\R)/
 SO(d-1,1) }$
to parameterise the coset space
or
$$
{
SL(d+1,\R)\over
 SO(d-1,2) }$$
In both cases, the expected discrete symmetry is
  $SL(d+1,\Z)$.

 For the bosonic string,   the dimensional reduction on a $d$ torus
 gives moduli in
$$
{
O(d,d)\over
O(d )\times O(d ) } \times \R
\eqn\abc$$
 or
$$
{
O(d,d)\over
O(d-1,1)\times O(d-1,1) } \times \R
\eqn\abc$$
together with
$2d$ vectors  from the reduction of the metric and 2-form gauge field.
In four dimensions S duality appears due to the 2-form,
and for compactifications
to 3 dimensions, the vectors are dual to scalars, and can be combined with
the other scalars to
take values in the coset spaces
$$
{
O(d+1,d+1)\over
O(d +1)\times O(d+1) }
\eqn\abc$$
 or
$$
{
O(d+1,d+1)\over
O(d-1,2)\times O(d-1,2) }
\eqn\abc$$
Thus the dilaton is combined with the other scalars and are acted on by
an $O(d+1,d+1;\Z)$ U-duality combining the $O(d ,d ;\Z)$ T-duality with
discrete
$SL(2,\Z)$ Ehlers-type symmetries, and the $SL(2,\Z)$ S-dualities inherited
from 4-dimensions, as in
[\Sena]. For compactifications to 3 dimensions of the heterotic string (with
$d=7,n=16$), or for any
theory of half-maximal supergravity coupled to
$n$ vector multiplets, a similar analysis gives the coset spaces
$$
{
O(d+1,d+n+1)\over
O(d +1)\times O(d+n+1) }
\eqn\abc$$
for  compactification on a spacelike torus [\Sena], or
$$
{
O(d+1,d+n+1)\over
O(d-1,2)\times O( d+n-1,2) }
\eqn\abc$$
for  compactification on a Lorentzian torus [\Sena].
For the heterotic string,
the coset space is
$$
{
O(8,24)\over
O(6,2)\times O( 22,2) }
\eqn\ert$$
and there is   an
$ O(8,24;\Z)$ U-duality symmetry [\Sena,\HT].

For reductions to 2   dimensions, the situation is more complicated  [\jujh,
\brma, \junis, \beju, \Senb].  In the reduction of the  heterotic string to 1+1
dimensions,
one obtains scalars in
$$ {O(8,24)\over O(8) \times O(24)} \times \R
$$
coupled to 2-dimensional gravity (which includes a coupling $\int d^2x \sqrt g
R e^{-\phi}$ where $\phi$ is the dilaton)  [\Senb].
The classical   supergravity theory has an affine $O(8,24)$ symmetry  [\Senb]
in which the central charge acts through scale transformations, and
a discrete subgroup of this is conjectured to be the U-duality  symmetry of the
full string theory
   [\HT, \Senb].

For the reduction on a Lorentzian torus the situation is similar, with scalars
taking values in
$$ {O(8,24)\over O(7,1) \times O(23,1)} \times \R
\eqn\erta$$
coupled to gravity, with an affine $O(8,24)$ symmetry.
However, if one first reduces on a Lorentzian torus to three dimensions,
dualises the vectors to scalars so  that the scalar target space is \ert, then
reduces to two dimensions, one obtains a coset space
$$
{
O(8,24)\over
O(6,2)\times O( 22,2) } \times \R
\eqn\ertb$$
which is different from \erta. However, in two Euclidean dimensions, the dual
of a scalar $\phi$ with kinetic term $ (D\phi )^2$ is a scalar $\tilde \phi$
with kinetic term $ -(D \tilde\phi )^2$, and some of the scalars in the coset
space
\ertb\ can be dualised so that the resulting coset space is \erta.
Thus in two Euclidean dimensions, duality can be used to analytically continue
the coset space to another real form, as T-dualities for Euclidean world-sheets
can change the target space signature.

One can also consider reduction to 0 dimensions. For the
bosonic  string, this was considered in
[\moore]. The metric, anti-symmetric tensor and dilaton moduli
take values in
$$\R \times {O(26,26)
\over O(25,1)\times O(25,1)}
\eqn\this$$
However, if one first compactifies to 3 Euclidean dimensions on $T^{22,1}$,
dualises the vectors to obtain scalars in $O(23,23)/O(21,2)\times O(21,2)$,
then compactifies the remaing three dimensions, one
obtains a moduli space which is a different real form of \this, and going via 2
dimensions, duality transformations could give further real forms of \this.
Similarly, for the heterotic string, the moduli
take values in
$$\R \times {O(10,26)
\over O(9,1)\times O(25,1)}
\eqn\abc$$
and other real forms of this can be obtained by going via 2 or 3 dimensions and
performing duality transformations.

 \noindent {\caps\enspace  Compactifications of 11-Dimensional Supergravity and
M-Theory}
\vskip\headskip
%\section{Compactifications of 11-Dimensional Supergravity and M-Theory}

We now turn to the toroidal compactification of M-theory. Compactification on
the
Euclidean torus $T^d$ gave the coset spaces $G_d/H_d$ in table 1, and the
U-duality group $G_d(\Z)$.
Compactification on $T^{d-1,1}$ gives a moduli space which we will assume to be
a coset space
 $\ti G_d/ \ti H_d$. This must have the following properties.
(i) As
$T^{d-1,1}$ contains both $T^{d-1,0}$ and $T^{d-2,1}$,
$G_{d-1}/H_{d-1}\subset \ti G_d/ \ti H_d$ and
$\ti G_{d-1}/\ti H_{d-1}\subset \ti G_d/ \ti H_d$.
(ii) Since $
{
GL(d,\R)/
 SO(d-1,1) } $ is the moduli space of metrics on $T^{d-1,1}$,
$$
{
GL(d,\R)\over
 SO(d-1,1) } \subset \ti G_d/ \ti H_d$$
(iii)
The same    theory
emerges from the type IIB theory compactified on $T^{d-2,1}$, and this has
$O(d,d)$ T-duality and a
commuting $SL(2)$ from the type IIB S-duality. Thus,
$${
O(d,d)\over
O(d-1,1)\times O(d-1,1) }\times {SL(2,\R)
\over U(1)}
\subset \ti G_d/ \ti H_d$$
(iv)
The dimension of $ \ti G_d/ \ti H_d$ is
the same as that of  $ G_d/H_d$.
It is parameterised
by constant Lorentzian metrics   $g_{ij}$ taking values in ${
GL(d,\R)/
 SO(d-1,1) } $
and an anti-symmetric tensor $A_{ijk}$ on $T^{d-1,1}$, together
with extra scalars from dualising anti-symmetric tensor gauge fields.
For the Euclidean torus,    $g_{ij}$ and $A_{ijk}$
parameterise
$${
GL(d,\R)\over
 SO(d) } \sd \R^{c_d}$$ where $c_d=d!/6(d-3)!$, while for $T^{d-1,1}$, the
$A_{ijk}$
split into    $A_{abc}$ with space-like norm and   $A_{ab0}$ with time-like
norm, so that
$g_{ij}$ and $A_{ijk}$
parameterise $${
GL(d,\R)\over
SO(d-1,1)  } \sd \R^{c_{d-1},t}$$ where $t=(d-1)(d-2)/2$  (recall   that
$\R^{c_{d},t} $ has $c_d$ spacelike dimensions
and
$t$ timelike ones). Thus
$${
GL(d,\R)\over
SO(d-1,1)  } \sd \R^{c_{d-1},t}\subseteq \ti G_d/ \ti H_d$$
In $D=5$, there is an extra scalar  that arises from dualising
$A_{\mu\nu\rho}$, in $D=4$ there are 7 scalars from dualising $A_{\mu\nu i}$
and in $D=3$ there are $28+8$  scalars from dualising $A_{\mu ij}$ and $g_{\mu
i}$.

The unique coset spaces $ \ti G_d/ \ti H_d$ satisfying these conditions
have $\ti G_d=G_d$ and $\ti H_d$ is a non-compact form of $H_d$. These
non-compact forms are given
in table 2.

\vskip 1cm

\begintable
D=11-d |
$G_d$ | $H_d$  | $\ti H_d$ |$ K_d $ \elt
 $10$ | $SO(1,1) $ | $\II $ | $ \II $ | $\II $ \elt
 $9$ | $SL(2,\R)\times SO(1,1)$ | $SO(2) $ | $SO(1,1)$  | $\II $ \elt
 $8$ | $SL(3,\R)\times SL(2,\R)$ |  $SO(3)\times SO(2) $ | $SO(2,1)\times
SO(2)$
| $SO(2)\times SO(2)$
\elt
 $7$ | $SL(5,\R)$ | $SO(5)$ | $SO(3,2)$ |  $SO(3)\times SO(2) $
\elt
 $6$ | $SO(5,5)$ | $SO(5)\times SO(5)$ |  $SO(5,C)$  | $SO(5)$ \elt
 $5$ | $E_{6(6)}$ | $USp(8)$ |  $USp(4,4)$ | $SO(5)\times SO(5)$ \elt
 $4$ | $E_{7(7)}$ | $SU(8)$ | $SU^*(8)$ | $USp(8)$  \elt
 $3$ | $E_{8(8)}$ | $SO(16)$ | $SO^*(16)$ |$U(8)$
 \elt
 $2$ | $E_{9(9)}$ | $H_2$ |$H_2^*$|$K_2$
\endtable

%\centerline
{{\bf Table 2} Toroidal reductions of M-theory   to
$D=11-d$.  Reduction on  $T^d$ gives the scalar  coset space  $G_d/H_d$
while reduction on  $T^{d-1,1}$ gives the scalar  coset space  $G_d/\ti H_d$.
$\ti H_d$ is a non-compact form of $H_d$ with maximal compact subgroup $K_d$.}

\vskip .5cm

$\ti H_d$ is a non-compact form of $H_d$ with maximal compact subgroup $K_d$,
so that the generators
of
$H_d$ can be decomposed into generators $k$ of $K_d$ and generators $h$ of
$H_d \setminus K_d$, and  the Lie
algebra takes the form
$$
[k,k]\sim k, \qquad [k,h]\sim h, \qquad
[h,h]\sim k
\eqn\abc$$
and $h,k$ can be taken to be anti-hermitian.
The algebra $\ti H_d$ is obtained from this by substituting $h \to \ti h  =
ih$, so that
the generators   $\ti h$ in $\ti H_d \setminus K_d$ are hermitian and so are
non-compact generators.

The bosonic part of the supergravity action in 3 dimensions is gravity
coupled to an $E_{8(8)}/SO^*(16)$ sigma-model and reducing to 2 dimensions
gives
scalars in $E_{8(8)}/SO^*(16)\times \R$ coupled to gravity,
with
an $E_{9(9)}$ symmetry.
In two dimensions,   $H_2^*$ is an infinite dimensional subgroup of  $E_{9(9)}$
containing
$SO^*(16)$,
which is a  fixed set of an   involution of $E_{9(9)}$.

\ack
{CMH would like to thank the Ecole Normale for hospitality.}

\refout

\bye